\documentclass[aps,pre,reprint,groupedaddress]{revtex4-1}

\usepackage{silence}
\usepackage{amssymb}
\usepackage{amsmath}
\usepackage{amsfonts}
\usepackage{graphicx}
\usepackage{soul}
\usepackage{float}

\usepackage{color}

\makeatletter

\newcommand{\Rmnum}[1]{\expandafter\@slowromancap\romannumeral #1@}
\makeatother

\begin{document}


\title{Spatial Control of Localized Oscillations in Arrays of Coupled Laser Dimers}


\author{Joniald Shena}
\email[]{jonialdshena@misis.ru}
\affiliation{National University of Science and Technology MISiS, Leninsky prosp. 4, Moscow, 119049, Russia}

\author{Yannis Kominis}
\affiliation{School of Applied Mathematical and Physical Science,
National Technical University of Athens, Athens, Greece}
\author{Anastasios Bountis}
\affiliation{Department of Mathematics, School of Science and Technology,
Nazarbayev University, Astana, Republic of Kazakhstan}
\author{Vassilios Kovanis}
\affiliation{Bradley Department of Electrical and Computer Engineering, Northern Virginia}


\date{\today}

\begin{abstract}

Arrays of coupled semiconductor lasers are systems possessing radically complex dynamics that makes them useful for numerous applications in beam forming and beam shaping. In this work, we investigate the spatial controllability of oscillation amplitudes in an array of coupled photonic dimers, each consisting of two semiconductor lasers driven by differential pumping rates. We consider parameter values for which each dimer's stable phase-locked state has become unstable through a Hopf bifurcation and we show that, by assigning appropriate pumping rate values to each dimer, large-amplitude oscillations coexist with negligibly small amplitude oscillations. The spatial profile of the amplitude of oscillations across the array can be dynamically controlled by appropriate pumping rate values in each dimer. This feature is shown to be quite robust, even for random detuning between the lasers, and suggests a mechanism for dynamically reconfigurable production of a large diversity of spatial profiles of laser amplitude oscillations.

\end{abstract}

\pacs{}
\keywords{semiconductor lasers \sep patterns controlling \sep nonlocal interaction \sep gain and loss \sep coupled array of lasers}

\maketitle

\section{INTRODUCTION}
\label{sec:introduction}
Laser arrays constitute a large family of nonlinear coupled systems which can exhibit a wide variety of complex dynamical behaviors. Although the emission from an individual unit is often stable, the coupled system can display synchronization, spatiotemporal collective phenomena \citep{Rogister2007,Chabanol2001,Shena2018} and also chaotic signals. \citep{Wang1988,
Herbert1990, Fabiny1993, Thornburg1997}. In recent years, there have been many studies concerning semiconductor lasers and their synchronization properties including the occurrence of chimera states \citep{Winful1992,Bohm2015,Shena2017,Shena2017a}. Moreover, there is an intense research interest in reconfigurable, multifunctional photonic devices that can be dynamically controlled. 

Photonic devices consisting of coupled lasers are non-Hermitian systems that are known to have unique mathematical and physical properties that have no counterpart in Hermitian ones and possess great potential for technological applications in integrated photonics and active metasurfaces.  A pair of coupled lasers constitutes a fundamental element for non-Hermitian optics which has recently been the subject of many investigations. 

In this direction, the consideration of differentially pumped coupled lasers has revealed a new set of interesting dynamical features such as localized synchronization \citep{Kovanis97}, tunable oscillations \citep{KominisarXiv}, existence of stable asymmetric phase-locked states \citep{KominisAsymmetric},  exceptional points and spectral transitions \citep{Kominis2017PT, KominisExceptional}, as well as tailored modulation response \citep{KominisResonance}. In these studies, the asymmetry is due to inhomogeneous pumping as well as frequency detuning and results in carrier densities above and below threshold, and with gain and loss coefficients of opposite signs in each laser, so that the respective electric fields experience varying gain and loss. It has been shown that stable asymmetric states exist even in absolutely symmetric configurations and that states of arbitrary asymmetry can be achieved by appropriate selection of the detuning and pumping profiles. Beam forming and steering applications are also based on differentially pumped coupled lasers and have been recently investigated both theoretically and experimentaly \citep{Choquette13, Choquette15}. Therefore, the role of the current injection provides an efficient mechanism for the control of the phase-locked and the oscillatory states of the system under a rich set of alternative options in the parameter values of the system that facilitate practical applications.

In the present work, we investigate the collective behavior of a large array of evanescently coupled semiconductor dimers with nearest-neighbor interactions. The crucial parameters for the control of the dynamical features are the coupling, the optical frequency detuning between the lasers and the pumping rates. Among them, only pumping is dynamically controllable and can be used to control the spatial profile of the oscillation amplitudes of the electric fields across the array. We find that by applying appropriate pumping schemes, characterized by different values in different groups of lasers we can ensure the coexistence of large amplitude oscillations for specific groups compared with negligibly small amplitude oscillations of other groups. Therefore, we show that we can dynamically generate different localization patterns in the same array, suggesting a reconfigurable mechanism with potential applications in multifunctional photonic devices.

\begin{figure*}[t]
\includegraphics[scale=0.5]{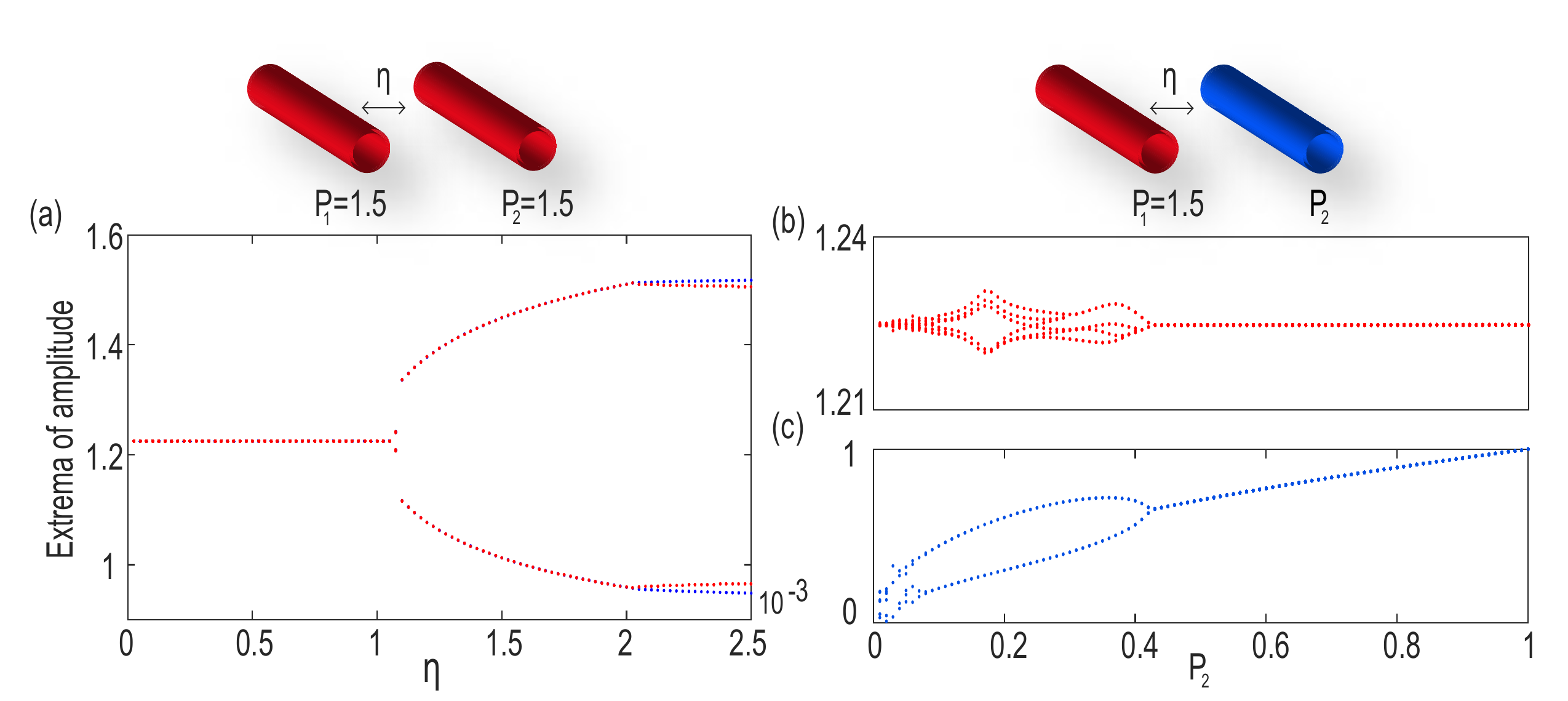}
\caption{Bifurcation diagrams for a single laser dimer. (a) Dependence of the extrema of the electric field amplitudes on the coupling strength for two identical lasers with $P_{1}=P_{2}=1.5$. The red color refers to the first laser and the blue to the second one. The steady state is shown to undergo a Hopf bifurcation at $\eta=0.00113$. Beyond the Hopf point, we observe different amplitude extrema for each laser. (b), (c) Extrema of the amplitudes of the electric fields as the second laser's pumping rate is varied, at constant coupling strength $\eta=0.0005$ and $P_{1}=1.5$. The label (b) refers to the first laser and (c) to the second one. The steady state, undergoes a reverse Hopf bifurcation at $P_{2}=0.42$. The asymmetry of the pumping rates drives the system to a stable limit cycle, with very small amplitudes of oscillation in (b) coexisting with much larger oscillation amplitudes shown in (c). Other parameters are $a=5$, $T=400$ and $\omega_{j}=0$. \label{fig_1}}
\end{figure*}

\begin{figure*}[t]
\includegraphics[scale=0.9]{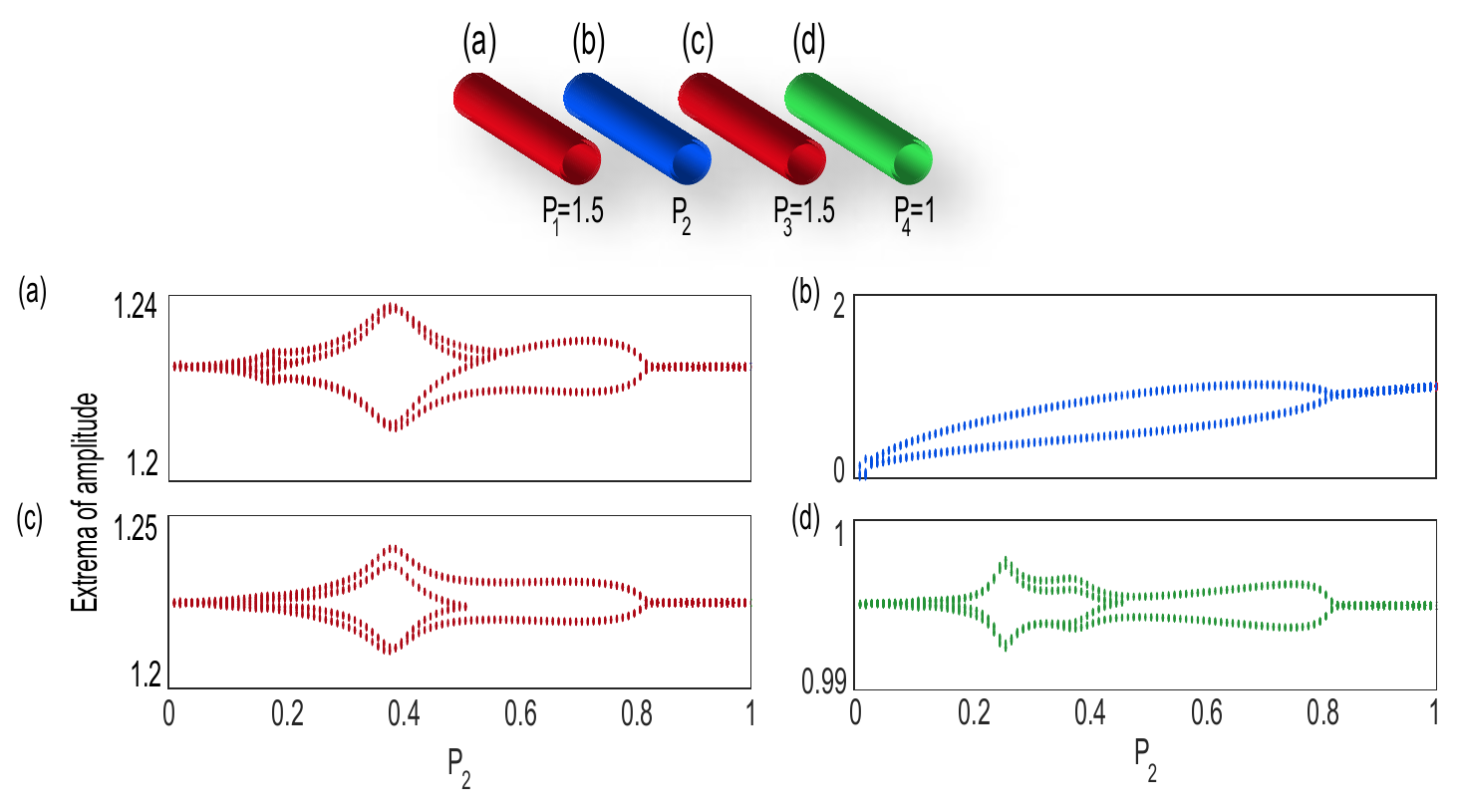}
\caption{Bifurcation diagrams for two coupled dimers. Dependence of extrema of the electric field amplitudes on the second laser's pumping rate for each laser, under inhomogeneous pumping. Label (a) refers to the first laser, (b) to the second, (c) to the third and (d) to the fourth. The phase locked state, undergoes a Hopf bifurcation at $P_{2} = 0.82$. All field amplitudes of the system are oscillating close to a fixed point, except for the second laser which exhibits much larger amplitude extrema. The pumping rates are $P_{1} = P_{3} = 1.5$ and $P_{4} = 1$. All other parameters are as in Fig. \ref{fig_1}}. \label{fig_2}
\end{figure*}

\section{THE MODEL}
\label{model}

The dynamics of an array of coupled laser dimers is governed by the following rate equations for the slowly varying complex amplitude $E_{j}$ of the electric field of each laser, and the corresponding population inversion $N_{j}$ \citep{Shena2017,Kominis2017,Parto2018}: 

\begin{subequations}
\begin{eqnarray}
\frac{d E_{j}}{dt} &=& (1-i a) N_{j} E_{j} \\ 
                   & & + i \eta(E_{j+1} + E_{j-1}) + i \omega_{j} E_{j} \nonumber \\
T \frac{dN_{j}}{dt} &=& P_{j} - N_{j} -(1+2 N_{j}) \lvert E_{j} \rvert^{2} \,.
\end{eqnarray}\label{eq1}  
\end{subequations}

\noindent where $ 1 \leq  j \leq  M $, $M$ being the number of lasers. The dimensionless time $t$ and the population inversion decay time $T$ are measured in units of the photon lifetime $\tau_{p}$.  $a$ is the linewidth enhancement factor resulting to amplitude to phase coupling. $P_{j}$ are the normalized pumping rates. $\eta$ is the normalized coupling rate between neighboring lasers due to interaction through their evanescent fields \citep{Zehnle2000}, and the normalized angular frequency $\omega_{j}$ measures the detuning of each laser from a common reference values.

Before proceeding to the investigation of the dynamics of a large array of coupled lasers, we consider the case of a single dimer to gain a better understanding of its dynamics. For equal pumping rates and in the absence of detuning, there exist two phase-locked states with  $0$ and $\pi$ phase differences between the two electric fields, which are stable for  $\eta>aP/(1+2P)$ and $\eta < (1+2P)/2aT$, respectively \citep{Winful1988}.

To understand the effect of the coupling strength, we plot in Fig. \ref{fig_1}(a) a bifurcation diagram of the maxima and minima of the amplitudes of the electric fields when the lasers have equal pumping rates set to $P=1.5$. Varying the coupling parameter, we see that a Hopf bifurcation to a stable limit cycle occurs at $\eta=0.00113$. As the coupling is further increased, the limit cycle breaks into a period-doubling sequence with different amplitude for each individual laser. This leads to a slightly asymmetric limit cycle bifurcating from a symmetric one, even under symmetric pumping \citep{KominisarXiv}. 

Next, keeping the coupling strength at the steady state value $\eta=0.0005$, we calculate in Fig. \ref{fig_1}(b) and (c) a bifurcation diagram, varying $P_{2}$ with $P_{1}=1.5$ fixed. The steady state is now seen to undergo a reverse Hopf bifurcation at $P_{2}=0.42$. Below this value, the first laser demonstrates multiple harmonics with amplitudes of different magnitude corresponding to the same limit cycle \cite{Varangis1997}. Remarkably, the oscillation amplitude of the first laser is much smaller than that of the second one, as  shown in Figs. \ref{fig_1} (b) and (c), respectively. Both lasers oscillate around mean values that correspond to a previously stable asymmetric phase-locked state. However, although the oscillations of the second laser have small mean values, they possess considerably larger peak-to-peak amplitude values in comparison to the first one. This behavior is absent in the case of equal pumping and is crucial for the study of a large array of coupled dimers carried out in the next section. The asymmetry characteristics as well as the frequencies of such limit cycles at the Hopf points have been recently studied in Y. Kominis et al \citep{KominisarXiv}. 

Following the same steps as in the above analysis, it is instructive to plot in Fig. \ref{fig_2} similar bifurcation diagrams for a system of two coupled dimers with different pumping rates ($P_{1}=P_{3}=1.5$ and $P_{4}=1$), keeping the coupling strength at the value $\eta=0.0005$ and varying $P_{2}$ as a control parameter. For low values of $P_{2}$ a limit cycle exists with multiple harmonics, until $P_{2}=0.82$, where the Hopf bifurcation occurs. Beyond that value, the system regains stability of the steady state and remains stable as $P_{2}$ continues to increase up to $P_{2} = 1$. Thus, through this choice of pumping parameters, it is shown that, in analogy to the previous case of a single dimer, the system as a whole supports an asymmetric localized oscillation where the second laser undergoes large oscillation amplitudes, whereas the other three oscillate with much smaller amplitudes for a large range of $P_{2}$. A generalization of this effect to the case of a large array of coupled dimers is studied in the next section.

\begin{figure}[t]
\includegraphics[scale=0.45]{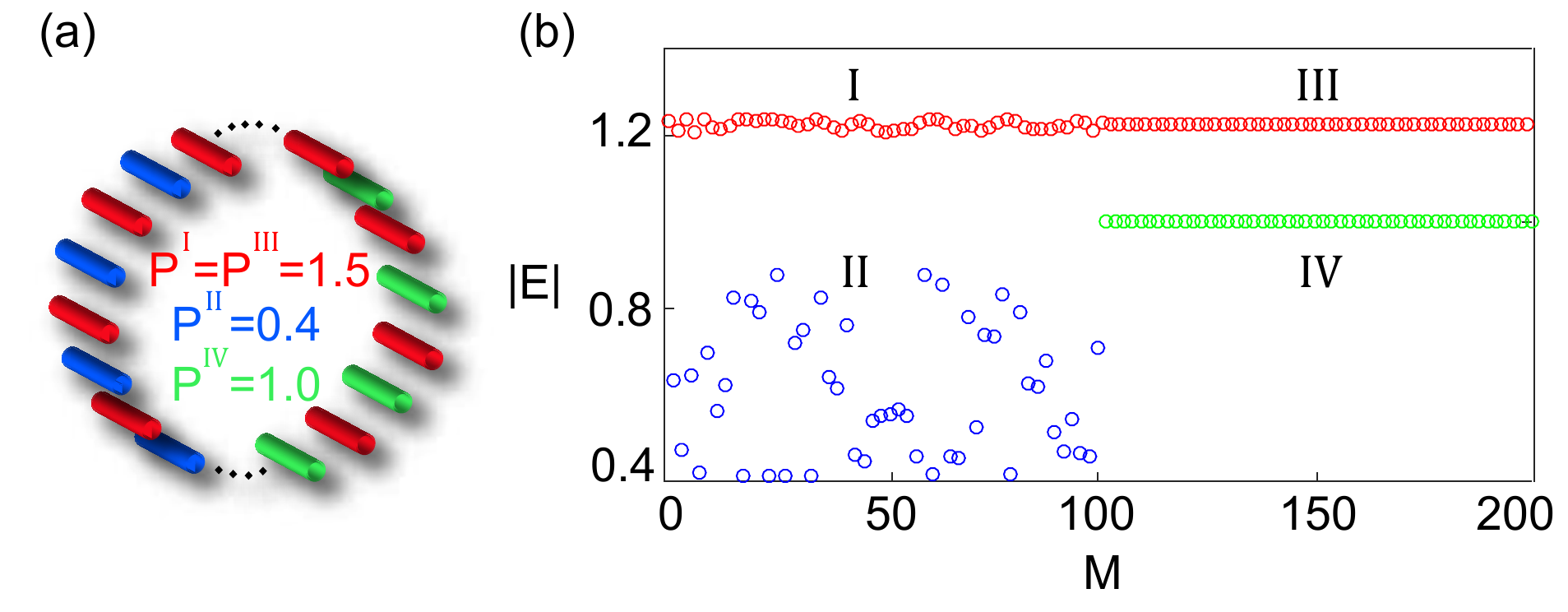}
\caption{(a) Schematic model of a circular array of 200 coupled lasers with inhomogeneous pumping. The array has been divided into two parts consisting of differently pumped dimers (four groups) where the dimers on the left are pumped with $P^{\Rmnum{1}}=1.5$ (red color, first group) and $P^{\Rmnum{2}}=0.4$ (blue color, second group) while the dimers of the right have $P^{\Rmnum{3}}=1.5$ (red color, third group) and $P^{\Rmnum{4}}=1$ (green color, fourth group). (b) Snapshots of the amplitude of the electric fields in an array of M=200 lasers with inhomogeneous pumping as in the model shown in (a). Only group \Rmnum{2} is shown to oscillate with large amplitudes while all others undergo small oscillations close to the fixed point. All other parameters are as in Fig. \ref{fig_1}.  \label{fig_3}}
\end{figure}

\begin{figure}[t]
\includegraphics[scale=0.55]{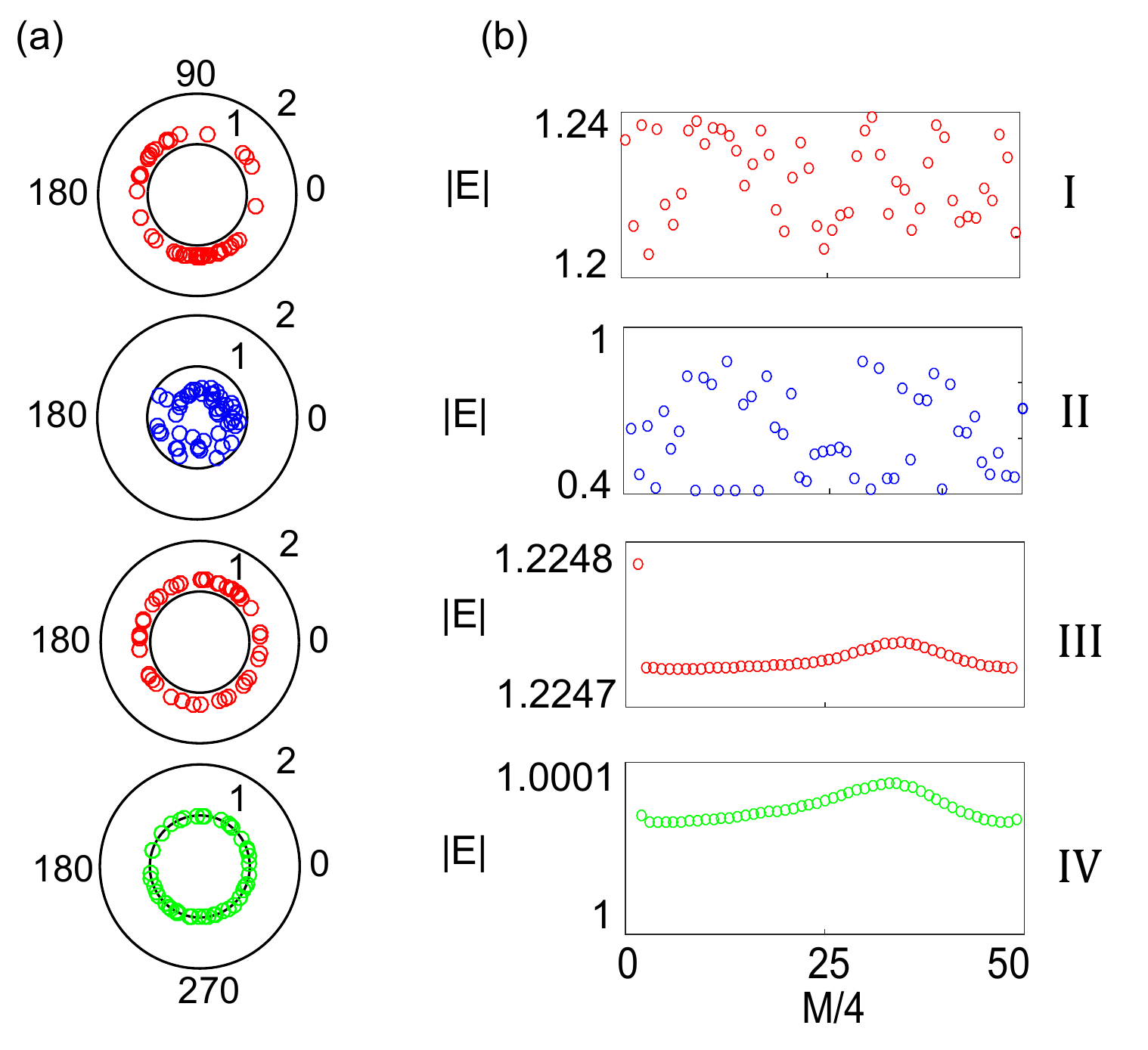}
\caption{ Snapshots of the complex electric fields distributions in the complex plane (a) and electric field amplitude (b), for all four groups as in Fig. \ref{fig_3}. 
\label{fig_4} }
\end{figure}
  
\section{AN ARRAY OF COUPLED LASER DIMERS} 

\begin{figure*}[t]
\includegraphics[scale=0.35]{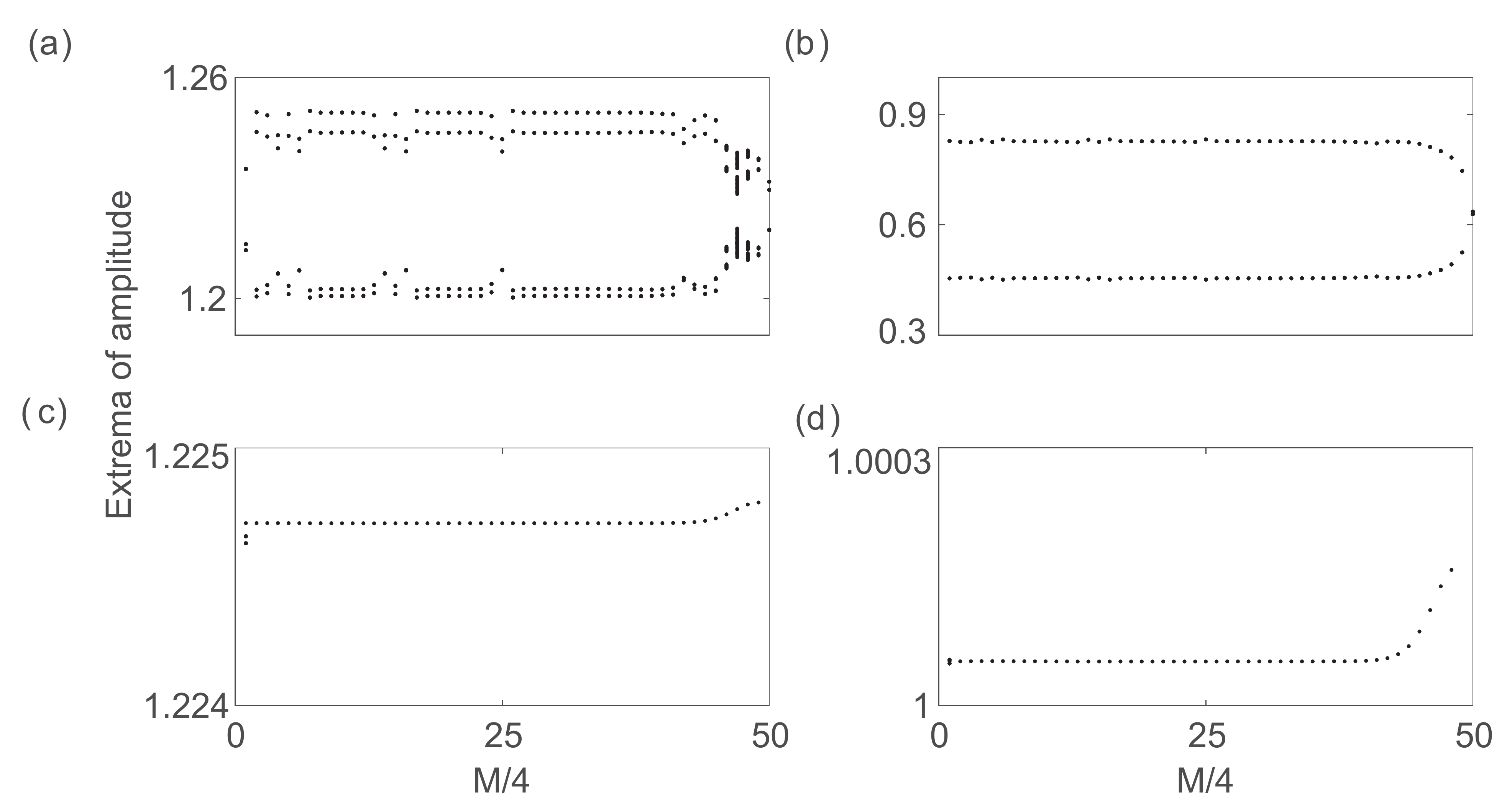}
\caption{Extrema of the electric field amplitudes for each laser of groups: (a) (I), (b) (II), (c) (III) and (d) (IV) as in Fig. \ref{fig_3} with the same parameters.   
\label{fig_41} }
\end{figure*}

\begin{figure}[t]
\includegraphics[scale=0.23]{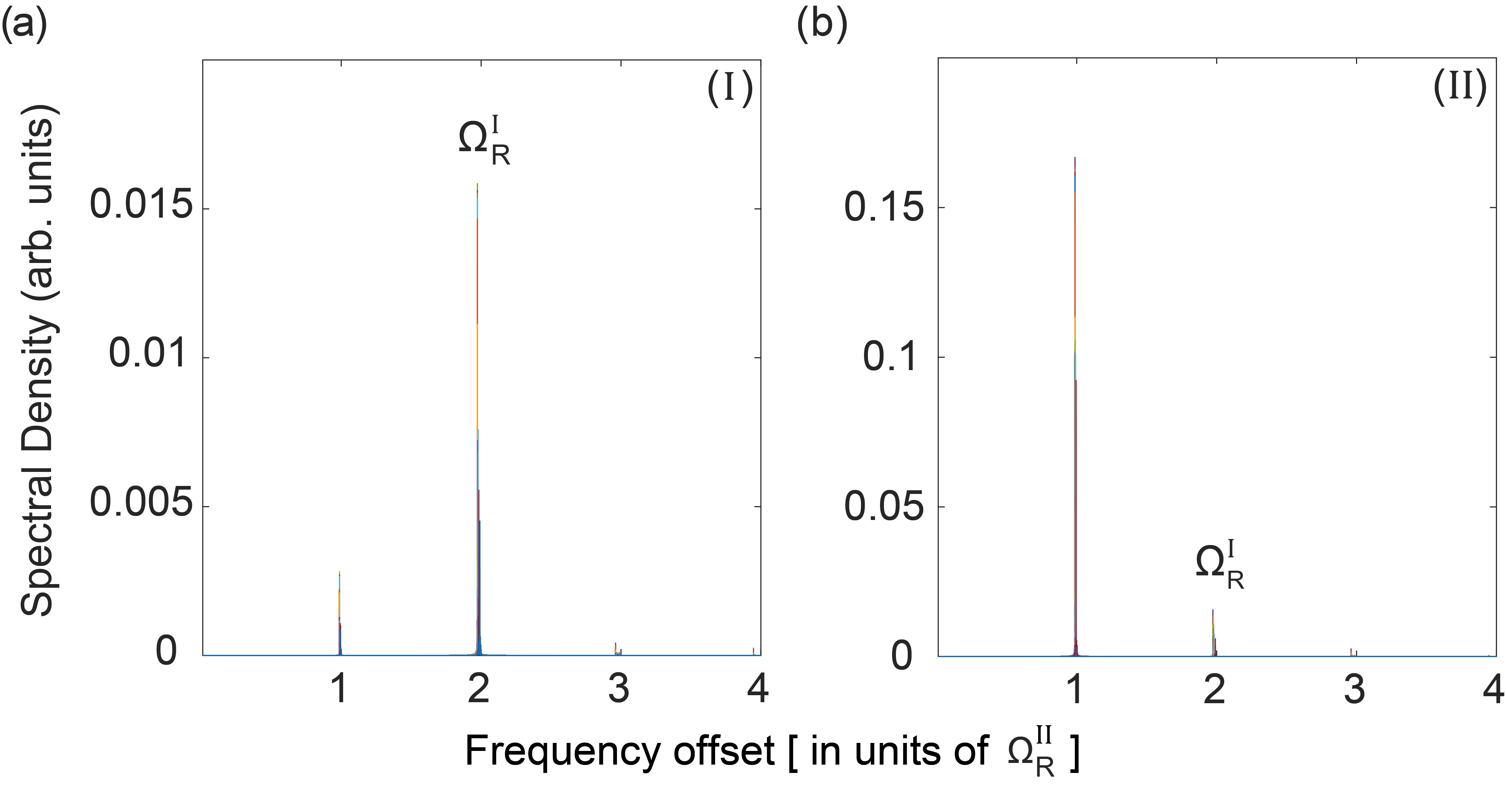}
\caption{ Fourier power spectra for all the lasers of groups (I) and (II) shown in Fig. \ref{fig_3}(b). The characteristic frequencies  $\Omega_{R}^{I}, \Omega_{R}^{II}$ are the normalized free-running laser relaxation oscillation frequencies of each individual laser of the groups (I) and (II), respectively, with $\Omega_{R}^{I} \simeq 2 \Omega_{R}^{II}$. 
\label{fig_42} }
\end{figure}

Let as now turn to a large circular array of coupled diode lasers, with nearest neighbor coupling and inhomogeneous pumping distribution, as shown in Fig. \ref{fig_3}(a). The network is divided in two clusters with two interfaces, or in four groups where the elements of the first group are pumped with rates $P_{j}^{\Rmnum{1}}=1.5$ for $j \in [1,3,5...M/2-1]$, the elements of the second group have $P^{\Rmnum{2}}=0.4$ for $j \in [2,4,6...M/2]$, in the third group $P^{\Rmnum{3}}=1.5$ for $j \in [M/2+1,M/2+3,...M-1]$, while in the fourth we set $P_{j}^{\Rmnum{4}}=1.0$ for $j \in [M/2+2,M/2+4,...M]$. To reveal its dynamics we integrate Eqs.(\ref{eq1}) using a fourth order Runge-Kutta algorithm, and employ  initial conditions with random phases taken from a uniform distribution over the interval [$-\pi$ to $\pi$], as well as random amplitudes and inverse populations. In all the numerical calculations we have performed in arrays consisting of 200 lasers, we have not observed any dependence on initial conditions. 

In Fig. \ref{fig_3}(b), we plot snapshots of the amplitude of the electric field for an array of M=200 lasers with inhomogeneous pumping distributed as in Fig. \ref{fig_3}(a), after a sufficiently long time $t=2*10^{5}$ to show the system's response after transient effects. The coupling strength between the neighbors is equal to $\eta=0.0005$. The present system represents an extension of the four laser model depicted in 
Fig. \ref{fig_2}. Under the applied distribution of pumping rates, the lasers of group \Rmnum{2} (blue color) support large amplitudes whereas the remaining groups demonstrate small amplitudes very close to the steady state. 

Snapshots of complex electric field distributions for every particular group are shown in Fig. \ref{fig_4}(a). The behavior is similar to that of the four coupled lasers where the second cluster attains large field amplitudes, while the other three oscillate close to the steady state. In Fig. \ref{fig_4}(b) we display the magnitudes and phases of these states depicted in the complex plane. Evidently, groups \Rmnum{1}, \Rmnum{3} and \Rmnum{4} exhibit small amplitude oscillations, whereas group \Rmnum{2} undergoes large amplitude oscillations. Groups \Rmnum{3}, \Rmnum{4} have a smooth amplitude distribution in sharp contrast to groups \Rmnum{1}, \Rmnum{2}. All groups have randomly distributed phases.

\begin{figure*}[t]
\includegraphics[scale=0.35]{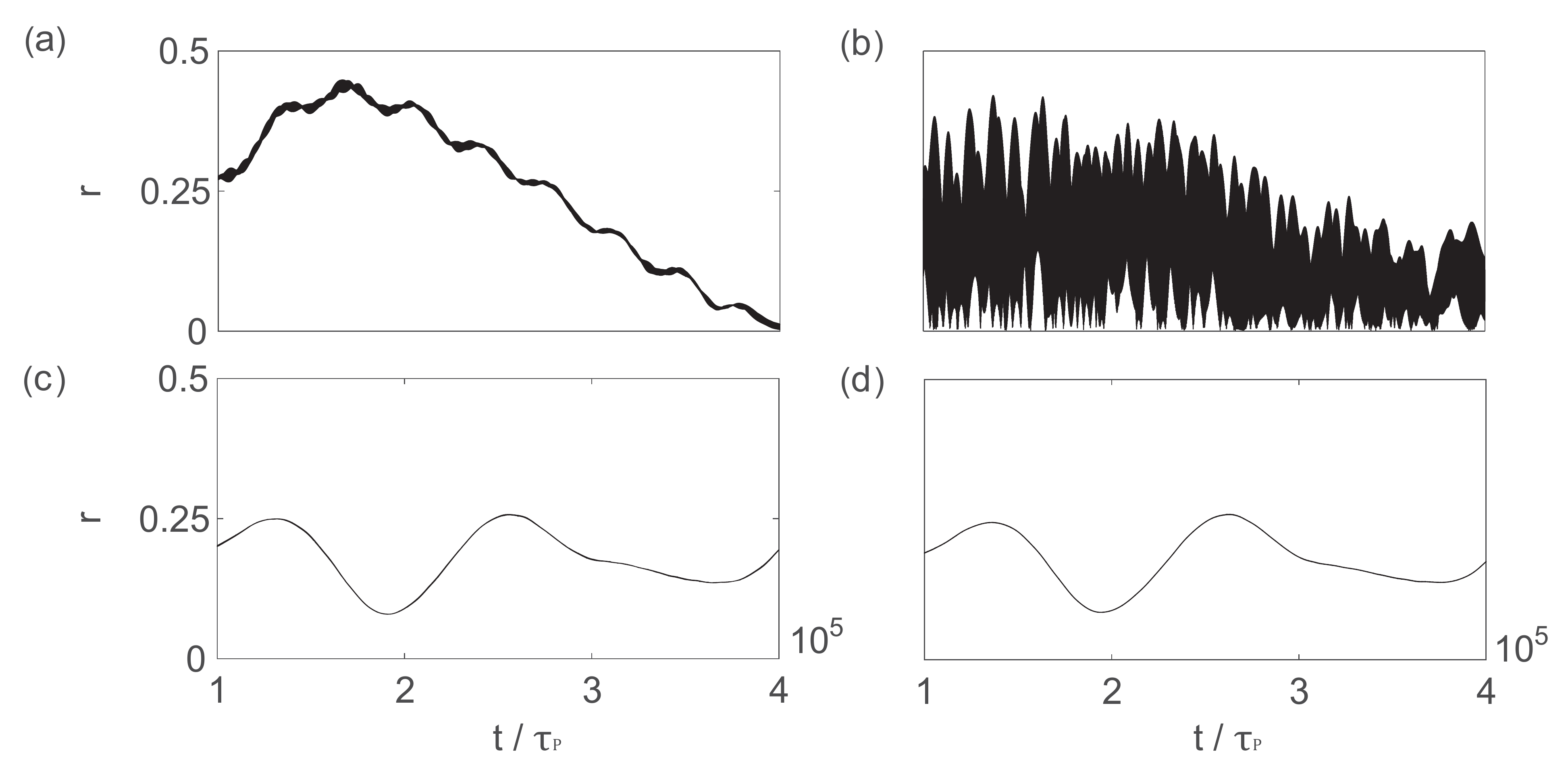}
\caption{ (a-d) The radius of the complex order parameter for laser oscillations in groups (I), (II), (III) and (IV) for a system as in Fig. \ref{fig_3}, in average over the time, indicating the lack of phase synchronization. 
\label{fig_43} }
\end{figure*}

\begin{figure}[t]
\includegraphics[scale=0.5]{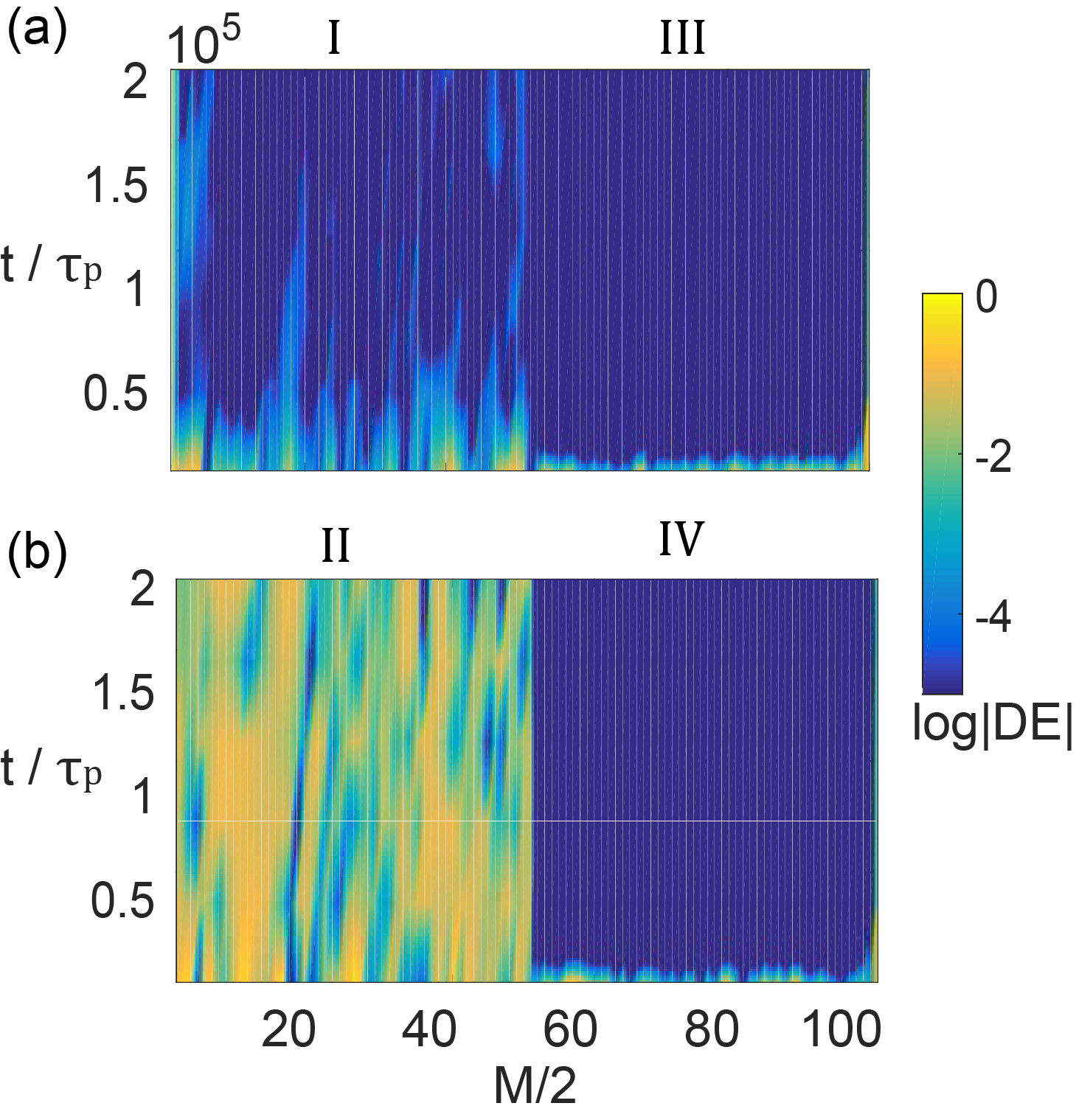}
\caption{Temporal evolution of the local curvature $ \lvert DE(t) \rvert $ in a logarithmic scale, for the array in Fig. \ref{fig_3} . For small oscillation amplitudes close to the steady state, as in groups \Rmnum{1}, \Rmnum{3} and \Rmnum{4}, the local curvature, properly rescaled with its max and min, is lower than $10^{-4}$, indicating very uniform amplitude distributions, in contrast with the large amplitudes case \Rmnum{2}, where the local curvature takes values close to unity. All the other parameters are as in Fig. \ref{fig_3}. \label{fig_5}}
\end{figure}

Extrema of the electric field amplitudes $|E|$ for each individual group (I), (II), (III) and (IV) seen in Fig. \ref{fig_3} and for the same parameters are shown in Fig. \ref{fig_41}(a), (b), (c) and (d) respectively. In the first group (I) (Fig. \ref{fig_41}(a))  we have double local extrema, except for the sites close to the second cluster, which are interacting with the groups (III) and (IV). The second group (II) (Fig. \ref{fig_41}(b)) undergoes oscillations with single maxima and minima. Finally, group (III) and (IV) (Fig. \ref{fig_41}(c) and (d)) undergo oscillations with very small amplitude that slightly increases for sites close to the first cluster which are interacting with the groups (I) and (II).  

Regarding the spectral content of the oscillations of each group, shown in  (Fig. \ref{fig_42}), both groups (I) and (II) have the same fundamental frequency which is close to the free-running relaxation frequency of the group (II), given as $\Omega_{R}^{II}= \sqrt{\frac{2P_{2}}{T}}$, and also have a second spectral peak close to $2\Omega_{R}^{II}$, which is also very close to $\Omega_{R}^{I} = 1.94 \Omega_{R}^{II}$ for the specific choice of pumping parameters. However, the relative power of the two spectral components is reversed in the two groups of lasers. 

In order to investigate the phase coherence of each group in time, we calculate the radius of the complex order parameter \citep{Kuramoto1984}
\begin{equation}
r(t) =  \frac{1}{N} \lvert \sum_{j=1}^{N} e^{i \phi_{j}} \rvert 
\end{equation}
where $N$ is the number of lasers in each group and $\phi_{j}$ the phase of the $j$th laser. 
If the group is coherent in phase, we have $r(t) \simeq 1$. On the other hand, if the group is incoherent, then $r(t) \simeq 0$. In Fig. \ref{fig_43}(a)-(d) we present the order parameter for groups (I), (II), (III) and (IV) oscillating as in Fig. \ref{fig_3} respectively. The oscillations of the order parameter for all the groups under $0.5$ indicate the lack of phase synchronization between the lasers of each group as also shown in the snapshot of Fig. \ref{fig_4}.

\begin{figure}[t]
\includegraphics[scale=0.5]{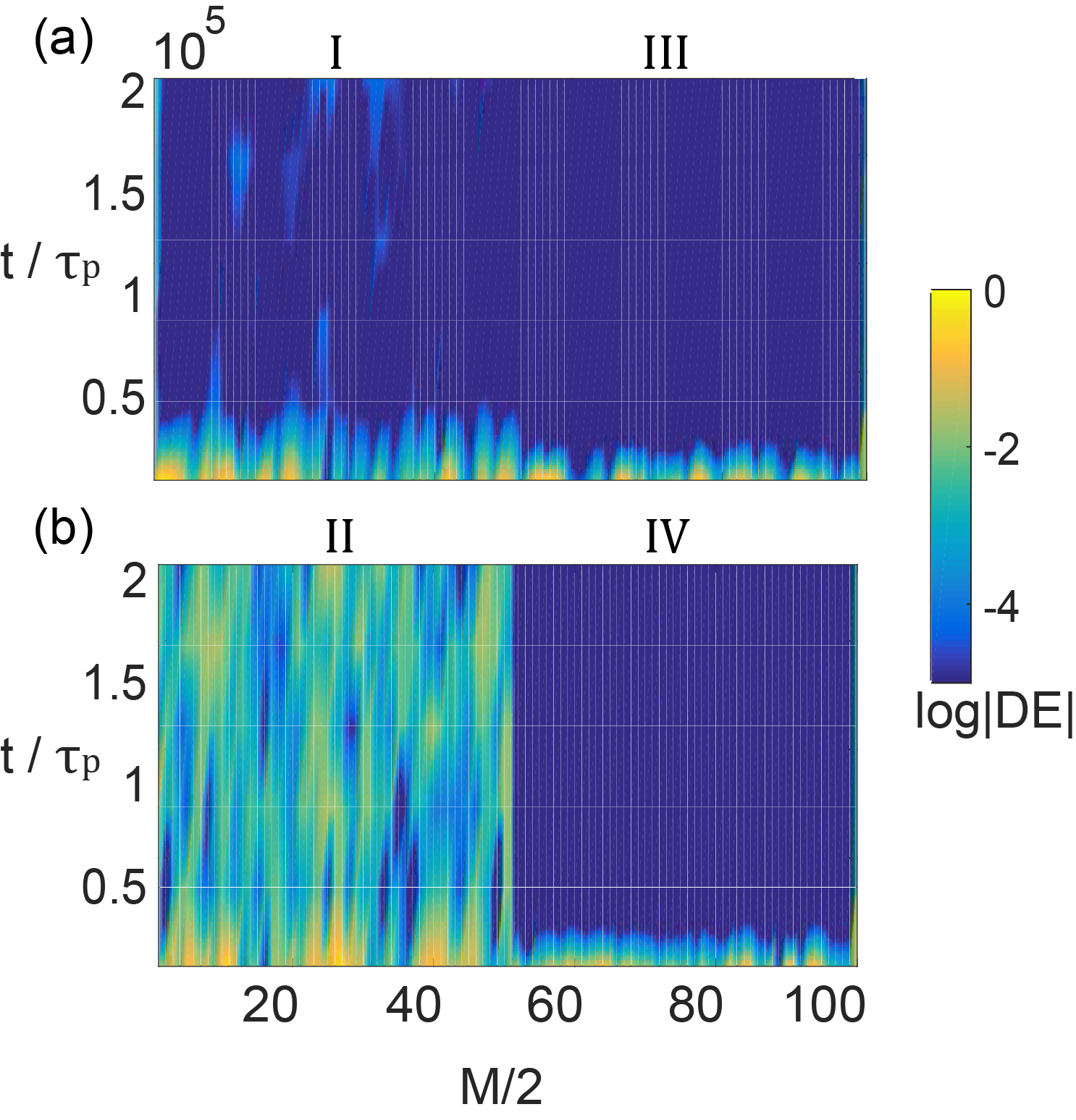}
\caption{Temporal evolution of the local curvature $\lvert DE(t) \rvert$ in a logarithmic scale as in Fig. \ref{fig_5}, in the presence of detuning. The detunings $\omega_{j}$ for all four groups are chosen from a normal random distribution with mean zero and standard deviation $\sigma_{\omega}=3\eta$. The behavior of the system is similar with that shown in Fig. \ref{fig_5}, implying that the coexistence of small and large oscillations remains robust even for a large values of detuning. All other parameters are as in Fig. \ref{fig_3}.  \label{fig_7}}
\end{figure}

To estimate the spatial coherence of the electric field amplitudes under time evolution, we  use, as an appropriate quantity, the local curvature of the electric field amplitude distribution within each laser group calculated for a sufficiently long time interval of $2*10^{5}$ \citep{Kemeth2016}. This is done by applying the discrete Laplacian $DE$ on the spatial data of the amplitude of the electric field as follows:

\begin{equation}
DE_{j}(t) = \lvert E_{j+1}(t) \rvert + \lvert E_{j-1}(t) \rvert - 2 \lvert E_{j}(t) \rvert
\end{equation}\label{eq4}

Figure \ref{fig_5} shows the spatio-temporal evolution of the absolute values of the local curvature in logarithmic scale corresponding to the oscillations shown in Fig. \ref{fig_3}(b). The local curvature, properly rescaled with respect to its maximum and minimum values, is shown to be significantly larger for group (II) in comparison to other laser groups. This particular behavior holds for any system size and suggests the possibility of achieving control of spatial patterns of large and small oscillations.

In all previous calculations, it has been assumed that the frequency detuning between all lasers is equal to zero. However, we have carried out an extensive study of what happens in the presence of detuning as follows: In Fig. \ref{fig_7}, the temporal evolution of the local curvature is presented for the electric field amplitudes shown in Fig. \ref{fig_3}(b), where the detunings $\omega_{j}$ have been chosen from a normal random distribution with mean zero and standard deviation equal to $\sigma_{\omega}=3\eta$. Note that, a proper consideration of the level of detuning, has to be made in relation to the coupling strength \citep{Adams2017,Gao2017}. In comparison to Fig. \ref{fig_5}, it is evident that all the qualitative characteristics of oscillation amplitude behaviors are quite robust even under the presence of strong random detuning.

\begin{figure}[t]
\includegraphics[scale=0.45]{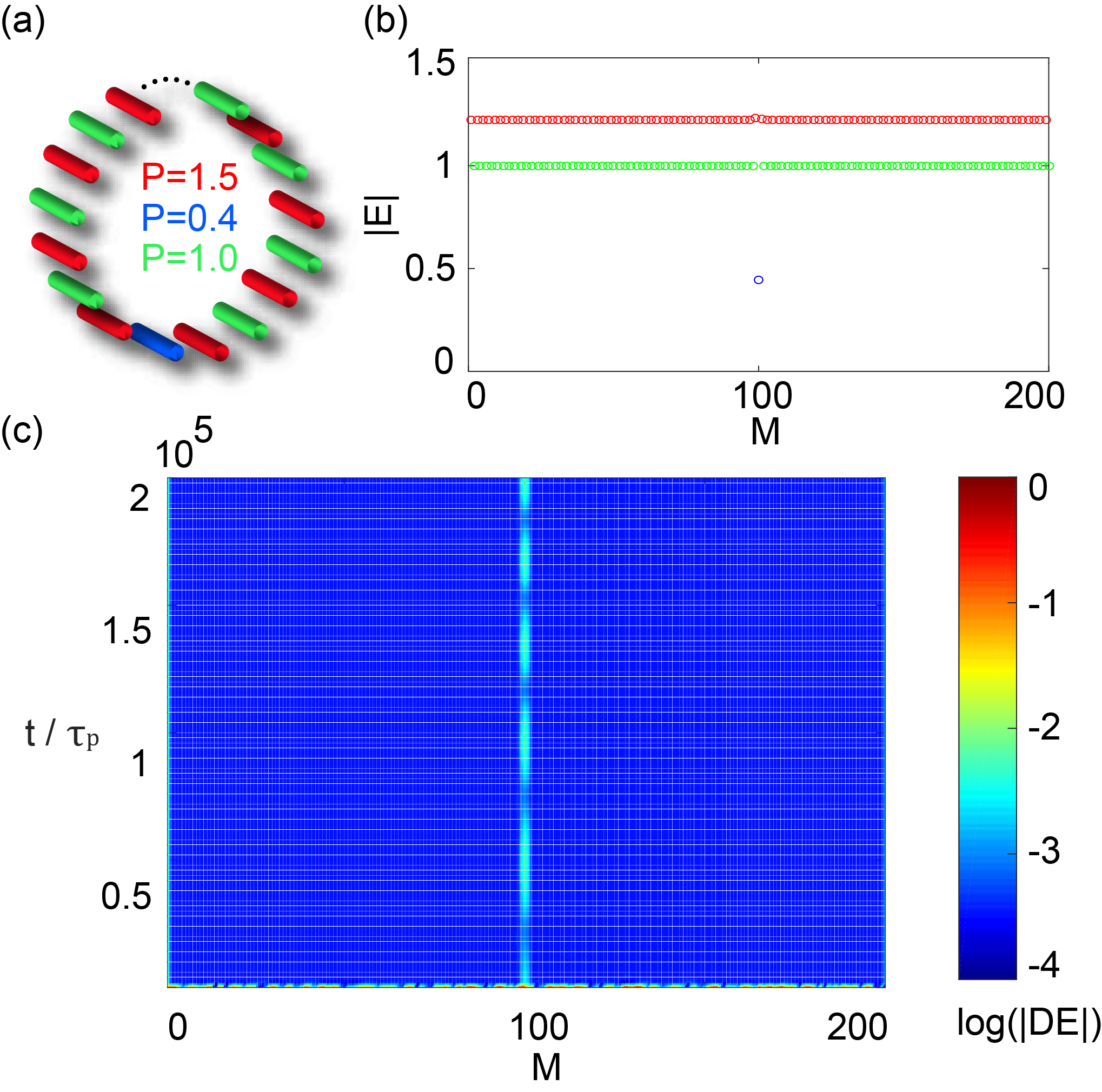}
\caption{(a) Schematic model for a circular array where the blue laser in the middle has $P = 0.4$, while the red ones have $P=1.5$  and the green ones $P =1.0$. (b) Snapshots of the amplitude of the electric field in an array of M=200 lasers with inhomogeneous pumping as in (a). (c) Temporal evolution of the local curvature $\lvert DE(t) \rvert$ in logarithmic scale, for the whole system. The coupling strength $\eta = 0.0005$ is sufficient for large oscillations of the dimer in the middle whereas the remaining part of the system exhibits oscillations of negligible amplitude. Other parameters: $a=5$, $T=400$ and $\omega_{j}=0$.  \label{fig_6}}
\end{figure}

As an example of extremely localized oscillations, we consider the case where only one laser of the array exhibits large amplitude oscillations, while all other lasers of the system are close to a steady state. Such a configuration is schematically presented in Fig. \ref{fig_6}(a), where all lasers are assigned pumping rate values alternating between $P=1.5$ and $P=1.0$ except the central laser which has $P=0.4$. Under this pumping distribution, Fig. \ref{fig_6}(b) displays a snapshot of the electric fields for the whole system, where the middle laser undergoes ``breather--like'' localized oscillations \citep{Mackay1994, Chen1996}. The temporal evolution of the local curvature corresponding to this configuration is plotted here in Fig. \ref{fig_6}(c). The coupling strength $\eta = 0.0005$ is sufficient to induce large amplitude oscillations for the middle laser, while the remaining part of the system is unaffected by the middle laser's oscillations.

One might conclude from the above simulations that the pumping rate can induce a kind of ``localization effect'' in the dynamics of the system. If, in a specific dimer, the first laser is pumped at a high enough rate compared with the second one, the latter can oscillate with large amplitudes without disturbing significantly the former, due the large pumping difference between them. On the other hand, if the pumping difference is small, both of them will oscillate with comparable amplitudes. Finally, if both of them have pumping rates in the regime preceding the Hopf bifurcation point, no oscillations are observed and the system remains in the steady state. This behavior appears to hold even for a large array of coupled dimers. More precisely, using appropriate coupling values, we can control the spatial pattern of amplitude oscillations spatial pattern of our dimers by controlling the distribution of pumping rates over the whole chain.


\section{CONCLUSIONS}\label{real}

In conclusion, we have shown that a priori selected spatially localized oscillations patterns can take place in a large array of active asymmetric photonic dimers, driven externally by judiciously choosing the pumping rates of the lasers in each dimer. The spatial profiles of the oscillation amplitudes can be dynamically controlled by pumping distributions to generate desired patterns for oscillation amplitudes of the electric fields of the lasers. 

More generally, we have shown that it is possible to achieve a wide variety of combinations of large amplitude oscillations with small field variations about unstable steady states, which remain localized for all time. However, even though the mechanism we have presented is capable of controlling the amplitudes, the phases of the electric fields remain, in general, uniformly distributed. The dynamical behavior of the system is also shown to be robust under different initial conditions, random frequency detuning and system size. 

Thus, we have demonstrated that spatial controllability of the oscillation amplitude profile of large arrays of coupled laser dimers can be achieved by appropriately selecting electrical pumping rates, which is the most conveniently accessible control parameter in chip scale diode systems. We believe that the controllability strategies presented in this work suggest a mechanism for multifunctional and reconfigurable photonic devices with multiple technological applications.

\section{ACKNOWLEDGEMENTS}
This work was supported by the Ministry of Education and Science of the Russian Federation in the framework of the Increase Competitiveness Program of NUST ``MISiS'' (Grant number K4-2018-049), the ORAU grant ``Taming Chimeras to Achieve the Superradiant Emitter'', 2017-2020, from Nazarbayev University and a grant from the Ministry of Education and Science of the Republic of Kazakhstan, via Contract number 059-2019.

\bibliography{bibliography}

\begin{thebibliography}{30}%
\makeatletter
\providecommand \@ifxundefined [1]{%
 \@ifx{#1\undefined}
}%
\providecommand \@ifnum [1]{%
 \ifnum #1\expandafter \@firstoftwo
 \else \expandafter \@secondoftwo
 \fi
}%
\providecommand \@ifx [1]{%
 \ifx #1\expandafter \@firstoftwo
 \else \expandafter \@secondoftwo
 \fi
}%
\providecommand \natexlab [1]{#1}%
\providecommand \enquote  [1]{``#1''}%
\providecommand \bibnamefont  [1]{#1}%
\providecommand \bibfnamefont [1]{#1}%
\providecommand \citenamefont [1]{#1}%
\providecommand \href@noop [0]{\@secondoftwo}%
\providecommand \href [0]{\begingroup \@sanitize@url \@href}%
\providecommand \@href[1]{\@@startlink{#1}\@@href}%
\providecommand \@@href[1]{\endgroup#1\@@endlink}%
\providecommand \@sanitize@url [0]{\catcode `\\12\catcode `\$12\catcode
  `\&12\catcode `\#12\catcode `\^12\catcode `\_12\catcode `\%12\relax}%
\providecommand \@@startlink[1]{}%
\providecommand \@@endlink[0]{}%
\providecommand \url  [0]{\begingroup\@sanitize@url \@url }%
\providecommand \@url [1]{\endgroup\@href {#1}{\urlprefix }}%
\providecommand \urlprefix  [0]{URL }%
\providecommand \Eprint [0]{\href }%
\providecommand \doibase [0]{http://dx.doi.org/}%
\providecommand \selectlanguage [0]{\@gobble}%
\providecommand \bibinfo  [0]{\@secondoftwo}%
\providecommand \bibfield  [0]{\@secondoftwo}%
\providecommand \translation [1]{[#1]}%
\providecommand \BibitemOpen [0]{}%
\providecommand \bibitemStop [0]{}%
\providecommand \bibitemNoStop [0]{.\EOS\space}%
\providecommand \EOS [0]{\spacefactor3000\relax}%
\providecommand \BibitemShut  [1]{\csname bibitem#1\endcsname}%
\let\auto@bib@innerbib\@empty
\bibitem [{\citenamefont {Rogister}\ and\ \citenamefont
  {Roy}(2007)}]{Rogister2007}%
  \BibitemOpen
  \bibfield  {author} {\bibinfo {author} {\bibfnamefont {F.}~\bibnamefont
  {Rogister}}\ and\ \bibinfo {author} {\bibfnamefont {R.}~\bibnamefont {Roy}},\
  }\href@noop {} {\bibfield  {journal} {\bibinfo  {journal} {Phys. Rev. Lett.}\
  }\textbf {\bibinfo {volume} {98}},\ \bibinfo {pages} {104101} (\bibinfo
  {year} {2007})}\BibitemShut {NoStop}%
\bibitem [{\citenamefont {Chabanol}\ and\ \citenamefont
  {Zehnl\'e}(2001)}]{Chabanol2001}%
  \BibitemOpen
  \bibfield  {author} {\bibinfo {author} {\bibfnamefont {M.}~\bibnamefont
  {Chabanol}}\ and\ \bibinfo {author} {\bibfnamefont {V.}~\bibnamefont
  {Zehnl\'e}},\ }\href@noop {} {\bibfield  {journal} {\bibinfo  {journal}
  {Phys. Rev. A}\ }\textbf {\bibinfo {volume} {63}},\ \bibinfo {pages} {053809}
  (\bibinfo {year} {2001})}\BibitemShut {NoStop}%
\bibitem [{\citenamefont {Shena}\ \emph {et~al.}(2018)\citenamefont {Shena},
  \citenamefont {Hizanidis}, \citenamefont {Kouvaris},\ and\ \citenamefont
  {Tsironis}}]{Shena2018}%
  \BibitemOpen
  \bibfield  {author} {\bibinfo {author} {\bibfnamefont {J.}~\bibnamefont
  {Shena}}, \bibinfo {author} {\bibfnamefont {J.}~\bibnamefont {Hizanidis}},
  \bibinfo {author} {\bibfnamefont {N.~E.}\ \bibnamefont {Kouvaris}}, \ and\
  \bibinfo {author} {\bibfnamefont {G.~P.}\ \bibnamefont {Tsironis}},\
  }\href@noop {} {\bibfield  {journal} {\bibinfo  {journal} {Phys. Rev. A}\
  }\textbf {\bibinfo {volume} {98}},\ \bibinfo {pages} {053817} (\bibinfo
  {year} {2018})}\BibitemShut {NoStop}%
\bibitem [{\citenamefont {Wang}\ and\ \citenamefont {Winful}(1988)}]{Wang1988}%
  \BibitemOpen
  \bibfield  {author} {\bibinfo {author} {\bibfnamefont {S.~S.}\ \bibnamefont
  {Wang}}\ and\ \bibinfo {author} {\bibfnamefont {H.~G.}\ \bibnamefont
  {Winful}},\ }\href@noop {} {\bibfield  {journal} {\bibinfo  {journal} {Appl.
  Phys. Lett}\ }\textbf {\bibinfo {volume} {52}},\ \bibinfo {pages} {1774}
  (\bibinfo {year} {1988})}\BibitemShut {NoStop}%
\bibitem [{\citenamefont {Winful}\ and\ \citenamefont
  {Rahman}(1990)}]{Herbert1990}%
  \BibitemOpen
  \bibfield  {author} {\bibinfo {author} {\bibfnamefont {H.~G.}\ \bibnamefont
  {Winful}}\ and\ \bibinfo {author} {\bibfnamefont {L.}~\bibnamefont
  {Rahman}},\ }\href@noop {} {\bibfield  {journal} {\bibinfo  {journal} {Phys.
  Rev. Lett.}\ }\textbf {\bibinfo {volume} {65}},\ \bibinfo {pages} {1575}
  (\bibinfo {year} {1990})}\BibitemShut {NoStop}%
\bibitem [{\citenamefont {Fabiny}\ \emph {et~al.}(1993)\citenamefont {Fabiny},
  \citenamefont {Colet}, \citenamefont {Roy},\ and\ \citenamefont
  {Lenstra}}]{Fabiny1993}%
  \BibitemOpen
  \bibfield  {author} {\bibinfo {author} {\bibfnamefont {L.}~\bibnamefont
  {Fabiny}}, \bibinfo {author} {\bibfnamefont {P.}~\bibnamefont {Colet}},
  \bibinfo {author} {\bibfnamefont {R.}~\bibnamefont {Roy}}, \ and\ \bibinfo
  {author} {\bibfnamefont {D.}~\bibnamefont {Lenstra}},\ }\href@noop {}
  {\bibfield  {journal} {\bibinfo  {journal} {Phys. Rev. A}\ }\textbf {\bibinfo
  {volume} {47}},\ \bibinfo {pages} {4287} (\bibinfo {year}
  {1993})}\BibitemShut {NoStop}%
\bibitem [{\citenamefont {Thornburg}\ \emph {et~al.}(1997)\citenamefont
  {Thornburg}, \citenamefont {M\"oller}, \citenamefont {Roy}, \citenamefont
  {Carr}, \citenamefont {Li},\ and\ \citenamefont {Erneux}}]{Thornburg1997}%
  \BibitemOpen
  \bibfield  {author} {\bibinfo {author} {\bibfnamefont {K.~S.}\ \bibnamefont
  {Thornburg}}, \bibinfo {author} {\bibfnamefont {M.}~\bibnamefont {M\"oller}},
  \bibinfo {author} {\bibfnamefont {R.}~\bibnamefont {Roy}}, \bibinfo {author}
  {\bibfnamefont {T.~W.}\ \bibnamefont {Carr}}, \bibinfo {author}
  {\bibfnamefont {R.-D.}\ \bibnamefont {Li}}, \ and\ \bibinfo {author}
  {\bibfnamefont {T.}~\bibnamefont {Erneux}},\ }\href@noop {} {\bibfield
  {journal} {\bibinfo  {journal} {Phys. Rev. E}\ }\textbf {\bibinfo {volume}
  {55}},\ \bibinfo {pages} {3865} (\bibinfo {year} {1997})}\BibitemShut
  {NoStop}%
\bibitem [{\citenamefont {Winful}(1992)}]{Winful1992}%
  \BibitemOpen
  \bibfield  {author} {\bibinfo {author} {\bibfnamefont {H.~G.}\ \bibnamefont
  {Winful}},\ }\href@noop {} {\bibfield  {journal} {\bibinfo  {journal} {Phys.
  Rev. A}\ }\textbf {\bibinfo {volume} {46}},\ \bibinfo {pages} {6093}
  (\bibinfo {year} {1992})}\BibitemShut {NoStop}%
\bibitem [{\citenamefont {B\"ohm}\ \emph {et~al.}(2015)\citenamefont {B\"ohm},
  \citenamefont {Zakharova}, \citenamefont {Sch\"oll},\ and\ \citenamefont
  {L\"udge}}]{Bohm2015}%
  \BibitemOpen
  \bibfield  {author} {\bibinfo {author} {\bibfnamefont {F.}~\bibnamefont
  {B\"ohm}}, \bibinfo {author} {\bibfnamefont {A.}~\bibnamefont {Zakharova}},
  \bibinfo {author} {\bibfnamefont {E.}~\bibnamefont {Sch\"oll}}, \ and\
  \bibinfo {author} {\bibfnamefont {K.}~\bibnamefont {L\"udge}},\ }\href@noop
  {} {\bibfield  {journal} {\bibinfo  {journal} {Phys. Rev. E}\ }\textbf
  {\bibinfo {volume} {91}},\ \bibinfo {pages} {040901} (\bibinfo {year}
  {2015})}\BibitemShut {NoStop}%
\bibitem [{\citenamefont {Shena}\ \emph
  {et~al.}(2017{\natexlab{a}})\citenamefont {Shena}, \citenamefont {Hizanidis},
  \citenamefont {Kovanis},\ and\ \citenamefont {Tsironis}}]{Shena2017}%
  \BibitemOpen
  \bibfield  {author} {\bibinfo {author} {\bibfnamefont {J.}~\bibnamefont
  {Shena}}, \bibinfo {author} {\bibfnamefont {J.}~\bibnamefont {Hizanidis}},
  \bibinfo {author} {\bibfnamefont {V.}~\bibnamefont {Kovanis}}, \ and\
  \bibinfo {author} {\bibfnamefont {G.~P.}\ \bibnamefont {Tsironis}},\
  }\href@noop {} {\bibfield  {journal} {\bibinfo  {journal} {Sci. Rep.}\
  }\textbf {\bibinfo {volume} {7}},\ \bibinfo {pages} {42116} (\bibinfo {year}
  {2017}{\natexlab{a}})}\BibitemShut {NoStop}%
\bibitem [{\citenamefont {Shena}\ \emph
  {et~al.}(2017{\natexlab{b}})\citenamefont {Shena}, \citenamefont {Hizanidis},
  \citenamefont {H\"ovel},\ and\ \citenamefont {Tsironis}}]{Shena2017a}%
  \BibitemOpen
  \bibfield  {author} {\bibinfo {author} {\bibfnamefont {J.}~\bibnamefont
  {Shena}}, \bibinfo {author} {\bibfnamefont {J.}~\bibnamefont {Hizanidis}},
  \bibinfo {author} {\bibfnamefont {P.}~\bibnamefont {H\"ovel}}, \ and\
  \bibinfo {author} {\bibfnamefont {G.~P.}\ \bibnamefont {Tsironis}},\
  }\href@noop {} {\bibfield  {journal} {\bibinfo  {journal} {Phys. Rev. E}\
  }\textbf {\bibinfo {volume} {96}},\ \bibinfo {pages} {032215} (\bibinfo
  {year} {2017}{\natexlab{b}})}\BibitemShut {NoStop}%
\bibitem [{\citenamefont {Hohl}\ \emph {et~al.}(1997)\citenamefont {Hohl},
  \citenamefont {Gavrielides}, \citenamefont {Erneux},\ and\ \citenamefont
  {Kovanis}}]{Kovanis97}%
  \BibitemOpen
  \bibfield  {author} {\bibinfo {author} {\bibfnamefont {A.}~\bibnamefont
  {Hohl}}, \bibinfo {author} {\bibfnamefont {A.}~\bibnamefont {Gavrielides}},
  \bibinfo {author} {\bibfnamefont {T.}~\bibnamefont {Erneux}}, \ and\ \bibinfo
  {author} {\bibfnamefont {V.}~\bibnamefont {Kovanis}},\ }\href@noop {}
  {\bibfield  {journal} {\bibinfo  {journal} {Phys. Rev. Lett.}\ }\textbf
  {\bibinfo {volume} {78}},\ \bibinfo {pages} {4745} (\bibinfo {year}
  {1997})}\BibitemShut {NoStop}%
\bibitem [{\citenamefont {Kominis}\ \emph
  {et~al.}(2019{\natexlab{a}})\citenamefont {Kominis}, \citenamefont
  {Bountis},\ and\ \citenamefont {Kovanis}}]{KominisarXiv}%
  \BibitemOpen
  \bibfield  {author} {\bibinfo {author} {\bibfnamefont {Y.}~\bibnamefont
  {Kominis}}, \bibinfo {author} {\bibfnamefont {A.}~\bibnamefont {Bountis}}, \
  and\ \bibinfo {author} {\bibfnamefont {V.}~\bibnamefont {Kovanis}},\
  }\href@noop {} {\bibfield  {journal} {\bibinfo  {journal} {arXiv:1911.04179}\
  } (\bibinfo {year} {2019}{\natexlab{a}})}\BibitemShut {NoStop}%
\bibitem [{\citenamefont {Kominis}\ \emph
  {et~al.}(2017{\natexlab{a}})\citenamefont {Kominis}, \citenamefont
  {Bountis},\ and\ \citenamefont {Kovanis}}]{KominisAsymmetric}%
  \BibitemOpen
  \bibfield  {author} {\bibinfo {author} {\bibfnamefont {Y.}~\bibnamefont
  {Kominis}}, \bibinfo {author} {\bibfnamefont {A.}~\bibnamefont {Bountis}}, \
  and\ \bibinfo {author} {\bibfnamefont {V.}~\bibnamefont {Kovanis}},\
  }\href@noop {} {\bibfield  {journal} {\bibinfo  {journal} {Phys. Rev. A}\
  }\textbf {\bibinfo {volume} {96}},\ \bibinfo {pages} {043836} (\bibinfo
  {year} {2017}{\natexlab{a}})}\BibitemShut {NoStop}%
\bibitem [{\citenamefont {Kominis}\ \emph
  {et~al.}(2017{\natexlab{b}})\citenamefont {Kominis}, \citenamefont
  {Kovanis},\ and\ \citenamefont {Bountis}}]{Kominis2017PT}%
  \BibitemOpen
  \bibfield  {author} {\bibinfo {author} {\bibfnamefont {Y.}~\bibnamefont
  {Kominis}}, \bibinfo {author} {\bibfnamefont {V.}~\bibnamefont {Kovanis}}, \
  and\ \bibinfo {author} {\bibfnamefont {T.}~\bibnamefont {Bountis}},\
  }\href@noop {} {\bibfield  {journal} {\bibinfo  {journal} {Phys. Rev. A}\
  }\textbf {\bibinfo {volume} {96}},\ \bibinfo {pages} {053837} (\bibinfo
  {year} {2017}{\natexlab{b}})}\BibitemShut {NoStop}%
\bibitem [{\citenamefont {Kominis}\ \emph {et~al.}(2018)\citenamefont
  {Kominis}, \citenamefont {Choquette}, \citenamefont {Kovanis},\ and\
  \citenamefont {Bountis}}]{KominisExceptional}%
  \BibitemOpen
  \bibfield  {author} {\bibinfo {author} {\bibfnamefont {Y.}~\bibnamefont
  {Kominis}}, \bibinfo {author} {\bibfnamefont {K.}~\bibnamefont {Choquette}},
  \bibinfo {author} {\bibfnamefont {V.}~\bibnamefont {Kovanis}}, \ and\
  \bibinfo {author} {\bibfnamefont {T.}~\bibnamefont {Bountis}},\ }\href@noop
  {} {\bibfield  {journal} {\bibinfo  {journal} {Appl. Phys. Lett}\ }\textbf
  {\bibinfo {volume} {113}},\ \bibinfo {pages} {081103} (\bibinfo {year}
  {2018})}\BibitemShut {NoStop}%
\bibitem [{\citenamefont {Kominis}\ \emph
  {et~al.}(2019{\natexlab{b}})\citenamefont {Kominis}, \citenamefont
  {Choquette}, \citenamefont {Kovanis},\ and\ \citenamefont
  {Bountis}}]{KominisResonance}%
  \BibitemOpen
  \bibfield  {author} {\bibinfo {author} {\bibfnamefont {Y.}~\bibnamefont
  {Kominis}}, \bibinfo {author} {\bibfnamefont {K.}~\bibnamefont {Choquette}},
  \bibinfo {author} {\bibfnamefont {V.}~\bibnamefont {Kovanis}}, \ and\
  \bibinfo {author} {\bibfnamefont {T.}~\bibnamefont {Bountis}},\ }\href@noop
  {} {\bibfield  {journal} {\bibinfo  {journal} {IEEE Photon. Journal}\
  }\textbf {\bibinfo {volume} {11}},\ \bibinfo {pages} {1500209} (\bibinfo
  {year} {2019}{\natexlab{b}})}\BibitemShut {NoStop}%
\bibitem [{\citenamefont {Johnson}\ \emph {et~al.}(2013)\citenamefont
  {Johnson}, \citenamefont {Siriani}, \citenamefont {Tan},\ and\ \citenamefont
  {Choquette}}]{Choquette13}%
  \BibitemOpen
  \bibfield  {author} {\bibinfo {author} {\bibfnamefont {M.}~\bibnamefont
  {Johnson}}, \bibinfo {author} {\bibfnamefont {D.}~\bibnamefont {Siriani}},
  \bibinfo {author} {\bibfnamefont {M.}~\bibnamefont {Tan}}, \ and\ \bibinfo
  {author} {\bibfnamefont {K.}~\bibnamefont {Choquette}},\ }\href@noop {}
  {\bibfield  {journal} {\bibinfo  {journal} {Appl. Phys. Lett.}\ }\textbf
  {\bibinfo {volume} {103}},\ \bibinfo {pages} {201115} (\bibinfo {year}
  {2013})}\BibitemShut {NoStop}%
\bibitem [{\citenamefont {Fryslie}\ \emph {et~al.}(2015)\citenamefont
  {Fryslie}, \citenamefont {Johnson},\ and\ \citenamefont
  {Choquette}}]{Choquette15}%
  \BibitemOpen
  \bibfield  {author} {\bibinfo {author} {\bibfnamefont {S.}~\bibnamefont
  {Fryslie}}, \bibinfo {author} {\bibfnamefont {M.}~\bibnamefont {Johnson}}, \
  and\ \bibinfo {author} {\bibfnamefont {K.}~\bibnamefont {Choquette}},\
  }\href@noop {} {\bibfield  {journal} {\bibinfo  {journal} {IEEE J. Quantum
  Electron.}\ }\textbf {\bibinfo {volume} {51}},\ \bibinfo {pages} {2600206}
  (\bibinfo {year} {2015})}\BibitemShut {NoStop}%
\bibitem [{\citenamefont {Kominis}\ \emph
  {et~al.}(2017{\natexlab{c}})\citenamefont {Kominis}, \citenamefont
  {Kovanis},\ and\ \citenamefont {Bountis}}]{Kominis2017}%
  \BibitemOpen
  \bibfield  {author} {\bibinfo {author} {\bibfnamefont {Y.}~\bibnamefont
  {Kominis}}, \bibinfo {author} {\bibfnamefont {V.}~\bibnamefont {Kovanis}}, \
  and\ \bibinfo {author} {\bibfnamefont {T.}~\bibnamefont {Bountis}},\
  }\href@noop {} {\bibfield  {journal} {\bibinfo  {journal} {Phys. Rev. A}\
  }\textbf {\bibinfo {volume} {96}},\ \bibinfo {pages} {043836} (\bibinfo
  {year} {2017}{\natexlab{c}})}\BibitemShut {NoStop}%
\bibitem [{\citenamefont {Parto}\ \emph {et~al.}(2018)\citenamefont {Parto},
  \citenamefont {Wittek}, \citenamefont {Hodaei}, \citenamefont {Harari},
  \citenamefont {Bandres}, \citenamefont {Ren}, \citenamefont {Rechtsman},
  \citenamefont {Segev}, \citenamefont {Christodoulides},\ and\ \citenamefont
  {Khajavikhan}}]{Parto2018}%
  \BibitemOpen
  \bibfield  {author} {\bibinfo {author} {\bibfnamefont {M.}~\bibnamefont
  {Parto}}, \bibinfo {author} {\bibfnamefont {S.}~\bibnamefont {Wittek}},
  \bibinfo {author} {\bibfnamefont {H.}~\bibnamefont {Hodaei}}, \bibinfo
  {author} {\bibfnamefont {G.}~\bibnamefont {Harari}}, \bibinfo {author}
  {\bibfnamefont {M.~A.}\ \bibnamefont {Bandres}}, \bibinfo {author}
  {\bibfnamefont {J.}~\bibnamefont {Ren}}, \bibinfo {author} {\bibfnamefont
  {M.~C.}\ \bibnamefont {Rechtsman}}, \bibinfo {author} {\bibfnamefont
  {M.}~\bibnamefont {Segev}}, \bibinfo {author} {\bibfnamefont {D.~N.}\
  \bibnamefont {Christodoulides}}, \ and\ \bibinfo {author} {\bibfnamefont
  {M.}~\bibnamefont {Khajavikhan}},\ }\href@noop {} {\bibfield  {journal}
  {\bibinfo  {journal} {Phys. Rev. Lett.}\ }\textbf {\bibinfo {volume} {120}},\
  \bibinfo {pages} {113901} (\bibinfo {year} {2018})}\BibitemShut {NoStop}%
\bibitem [{\citenamefont {Zehnl{\'{e}}}(2000)}]{Zehnle2000}%
  \BibitemOpen
  \bibfield  {author} {\bibinfo {author} {\bibfnamefont {V.}~\bibnamefont
  {Zehnl{\'{e}}}},\ }\href@noop {} {\bibfield  {journal} {\bibinfo  {journal}
  {Phys. Rev. A}\ }\textbf {\bibinfo {volume} {62}},\ \bibinfo {pages} {033814}
  (\bibinfo {year} {2000})}\BibitemShut {NoStop}%
\bibitem [{\citenamefont {Winful}\ and\ \citenamefont
  {Wang}(1988)}]{Winful1988}%
  \BibitemOpen
  \bibfield  {author} {\bibinfo {author} {\bibfnamefont {H.~G.}\ \bibnamefont
  {Winful}}\ and\ \bibinfo {author} {\bibfnamefont {S.~S.}\ \bibnamefont
  {Wang}},\ }\href@noop {} {\bibfield  {journal} {\bibinfo  {journal} {Appl.
  Phys. Lett}\ }\textbf {\bibinfo {volume} {53}},\ \bibinfo {pages} {1894}
  (\bibinfo {year} {1988})}\BibitemShut {NoStop}%
\bibitem [{\citenamefont {Varangis}\ \emph {et~al.}(1997)\citenamefont
  {Varangis}, \citenamefont {Gavrielides}, \citenamefont {Erneux},
  \citenamefont {Kovanis},\ and\ \citenamefont {Lester}}]{Varangis1997}%
  \BibitemOpen
  \bibfield  {author} {\bibinfo {author} {\bibfnamefont {P.~M.}\ \bibnamefont
  {Varangis}}, \bibinfo {author} {\bibfnamefont {A.}~\bibnamefont
  {Gavrielides}}, \bibinfo {author} {\bibfnamefont {T.}~\bibnamefont {Erneux}},
  \bibinfo {author} {\bibfnamefont {V.}~\bibnamefont {Kovanis}}, \ and\
  \bibinfo {author} {\bibfnamefont {L.~F.}\ \bibnamefont {Lester}},\
  }\href@noop {} {\bibfield  {journal} {\bibinfo  {journal} {Phys. Rev. Lett.}\
  }\textbf {\bibinfo {volume} {78}},\ \bibinfo {pages} {2353} (\bibinfo {year}
  {1997})}\BibitemShut {NoStop}%
\bibitem [{\citenamefont {Kuramoto}(1984)}]{Kuramoto1984}%
  \BibitemOpen
  \bibfield  {author} {\bibinfo {author} {\bibfnamefont {Y.}~\bibnamefont
  {Kuramoto}},\ }\href@noop {} {\emph {\bibinfo {title} {Chemical Oscillations,
  Waves, and Turbulence}}}\ (\bibinfo  {publisher} {Springer, Berlin,
  Heidelberg},\ \bibinfo {year} {1984})\BibitemShut {NoStop}%
\bibitem [{\citenamefont {Kemeth}\ \emph {et~al.}(2016)\citenamefont {Kemeth},
  \citenamefont {Haugland}, \citenamefont {Schmidt}, \citenamefont
  {Kevrekidis},\ and\ \citenamefont {Krischer}}]{Kemeth2016}%
  \BibitemOpen
  \bibfield  {author} {\bibinfo {author} {\bibfnamefont {F.~P.}\ \bibnamefont
  {Kemeth}}, \bibinfo {author} {\bibfnamefont {S.~W.}\ \bibnamefont
  {Haugland}}, \bibinfo {author} {\bibfnamefont {L.}~\bibnamefont {Schmidt}},
  \bibinfo {author} {\bibfnamefont {I.~G.}\ \bibnamefont {Kevrekidis}}, \ and\
  \bibinfo {author} {\bibfnamefont {K.}~\bibnamefont {Krischer}},\ }\href@noop
  {} {\bibfield  {journal} {\bibinfo  {journal} {Chaos}\ }\textbf {\bibinfo
  {volume} {26}},\ \bibinfo {pages} {094815} (\bibinfo {year}
  {2016})}\BibitemShut {NoStop}%
\bibitem [{\citenamefont {Adams}\ \emph {et~al.}(2017)\citenamefont {Adams},
  \citenamefont {Li}, \citenamefont {Cemlyn}, \citenamefont {Susanto},\ and\
  \citenamefont {Henning}}]{Adams2017}%
  \BibitemOpen
  \bibfield  {author} {\bibinfo {author} {\bibfnamefont {M.~J.}\ \bibnamefont
  {Adams}}, \bibinfo {author} {\bibfnamefont {N.}~\bibnamefont {Li}}, \bibinfo
  {author} {\bibfnamefont {B.~R.}\ \bibnamefont {Cemlyn}}, \bibinfo {author}
  {\bibfnamefont {H.}~\bibnamefont {Susanto}}, \ and\ \bibinfo {author}
  {\bibfnamefont {I.~D.}\ \bibnamefont {Henning}},\ }\href@noop {} {\bibfield
  {journal} {\bibinfo  {journal} {Phys. Rev. A}\ }\textbf {\bibinfo {volume}
  {95}},\ \bibinfo {pages} {053869} (\bibinfo {year} {2017})}\BibitemShut
  {NoStop}%
\bibitem [{\citenamefont {Gao}\ \emph {et~al.}(2017)\citenamefont {Gao},
  \citenamefont {Fryslie}, \citenamefont {Thompson}, \citenamefont {Carney},\
  and\ \citenamefont {Choquette}}]{Gao2017}%
  \BibitemOpen
  \bibfield  {author} {\bibinfo {author} {\bibfnamefont {Z.}~\bibnamefont
  {Gao}}, \bibinfo {author} {\bibfnamefont {S.~T.~M.}\ \bibnamefont {Fryslie}},
  \bibinfo {author} {\bibfnamefont {B.~J.}\ \bibnamefont {Thompson}}, \bibinfo
  {author} {\bibfnamefont {P.~S.}\ \bibnamefont {Carney}}, \ and\ \bibinfo
  {author} {\bibfnamefont {K.~D.}\ \bibnamefont {Choquette}},\ }\href@noop {}
  {\bibfield  {journal} {\bibinfo  {journal} {Optica}\ }\textbf {\bibinfo
  {volume} {4}},\ \bibinfo {pages} {323} (\bibinfo {year} {2017})}\BibitemShut
  {NoStop}%
\bibitem [{\citenamefont {MacKay}\ and\ \citenamefont
  {Aubry}(1994)}]{Mackay1994}%
  \BibitemOpen
  \bibfield  {author} {\bibinfo {author} {\bibfnamefont {R.~S.}\ \bibnamefont
  {MacKay}}\ and\ \bibinfo {author} {\bibfnamefont {S.}~\bibnamefont {Aubry}},\
  }\href@noop {} {\bibfield  {journal} {\bibinfo  {journal} {Nonlinearity}\
  }\textbf {\bibinfo {volume} {7}},\ \bibinfo {pages} {1623} (\bibinfo {year}
  {1994})}\BibitemShut {NoStop}%
\bibitem [{\citenamefont {Chen}\ \emph {et~al.}(1996)\citenamefont {Chen},
  \citenamefont {Aubry},\ and\ \citenamefont {Tsironis}}]{Chen1996}%
  \BibitemOpen
  \bibfield  {author} {\bibinfo {author} {\bibfnamefont {D.}~\bibnamefont
  {Chen}}, \bibinfo {author} {\bibfnamefont {S.}~\bibnamefont {Aubry}}, \ and\
  \bibinfo {author} {\bibfnamefont {G.~P.}\ \bibnamefont {Tsironis}},\
  }\href@noop {} {\bibfield  {journal} {\bibinfo  {journal} {Phys. Rev. Lett.}\
  }\textbf {\bibinfo {volume} {77}},\ \bibinfo {pages} {4776} (\bibinfo {year}
  {1996})}\BibitemShut {NoStop}%
\end{thebibliography}%

\end{document}